\documentclass[doublecol]{epl2} 

\usepackage{graphicx}
\usepackage{dcolumn}
\usepackage{bm}
\usepackage{amssymb}

\title{Soft Confinement for Polymer Solutions}
\shorttitle{Soft Confinement for Polymer Solutions}

\author{Yutaka Oya \and Toshihiro Kawakatsu}
\shortauthor{Y. Oya and T. Kawakatsu}

\institute{Department of Physics, Tohoku University, Sendai, 980-8578, Japan}

\abstract{
As a model of soft confinement for polymers, we investigated equilibrium shapes of a flexible vesicle that contains a phase-separating polymer solution.  
To simulate such a system, we combined the phase field theory (PFT) for the vesicle and the self-consistent field theory (SCFT) for the polymer solution.  
We observed a transition from a symmetric prolate shape of the vesicle to an asymmetric pear shape induced by the domain structure of the enclosed polymer solution.  
Moreover, when a non-zero spontaneous curvature of the vesicle is introduced, a re-entrant transition between the prolate and the dumbbell shapes of the vesicle is observed.  
This re-entrant transition is explained by considering the competition between the loss of conformational entropy and that of translational entropy of polymer chains due to the confinement by the deformable vesicle.  This finding is in accordance with the recent experimental result reported by Terasawa, {\it et al}.}
\begin{document}

\maketitle
Phase separated structures of polymer melts and polymer solutions induced by a confinement into a narrow space are actively investigated both experimentally and theoretically.  
For example, Wu {\it et al.} reported experimental observations of helical structures of diblock-copolymers induced by a confinement into a nano-sized cylindrical tube\cite{Wu}.  
In theoretical studies, self-consistent field theory (SCFT) is a powerful method to obtain the equilibrium structure of polymer solutions, because SCFT can take the polymer conformations into account\cite{Kawakatsu,Fredrickson}.  
In preceding studies, using the SCFT, effects of confinement on polymer melts and solutions have been investigated for various types of containers, such as those with spherical and cylindrical shapes\cite{Wu,Yu,Li,Sevink}.  
However, these studies are limited to the cases with hard confinements of polymer solutions by rigid containers.  

In the present article, we study a gsoft confinementh of polymer solutions, where the polymer solutions are enclosed by flexible containers.  
As a target system of this soft confinement, we study equilibrium structures of a flexible vesicle that encloses a polymer solution.  
Such polymer-containing vesicles can be frequently found in biological systems such as endocytosis and exocytosis, and are expected to be applicable to industrial science, for example the drug-delivery system.  
Recently, Terasawa {\it et al.} and Nakaya {\it et al.} showed shape deformations of vesicles\cite{Terasawa} or closed membranes\cite{Nakaya} induced by enclosed polymers.  
They suggested an important effect of the translational entropy of the enclosed polymers.  

In the present article, to study the deformation of polymer-containing vesicles theoretically, we apply our field-theoretic approach\cite{Oya} where SCFT for polymers\cite{Kawakatsu,Fredrickson} and phase field theory (PFT) for the vesicle shape\cite{Du,Biben,Campelo} are combined.  In our previous publication\cite{Oya}, we discussed a transition of a polymer-containing vesicle between a prolate shape and an oblate shape, where we limited our discussions to the cases with polymers swollen by a good solvent and vesicles that have no spontaneous curvature.  However, in realistic situations, enclosed polymers are often in a globular state ({\it i.e.} the solvent is a poor solvent) and the vesicle has non-zero spontaneous curvature due to the asymmetric composition between the inner and outer leaflets of the bilayer membrane.  In the present article, we will show remarkable effects of these two features on the vesicle deformation.
 
Let us describe our theoretical model for polymer-containing vesicles. In our model, the vesicle is modeled by PFT, where a scalar field $\psi({\bf r})$ which is called the "phase field"\cite{Du,Biben,Campelo} specifies the inside and the outside regions of the vesicle by its positive and negative regions.
Using an analogy to the Ginzburg-Landau theory for phase separating binary mixtures, the density distribution of the surfactant molecules that compose the vesicle, denoted as $\phi_{\rm m}({\bf r})$, and the total surface area of the vesicle ${\cal S}$ are represented in terms of the phase field
\begin{eqnarray}
\label{surface_1}
\phi_{\rm m} ({\bf r}) &=& \frac{1}{2} (1 - \psi({\bf r})^{2})^{2} + \epsilon^{2} |\nabla \psi({\bf r})|^{2}, \\
\label{surface_2}
{\cal S}              &=& \frac{3 \sqrt{2}}{8 \epsilon} \int \phi_{\rm m} ({\bf r}) d{\bf r}.
\end{eqnarray}
Equation (\ref{surface_2}) corresponds to the interfacial energy of the Ginzburg-Landau model, which is proportional to the total interfacial area.  

The shape of the vesicle is characterized by Helfrich's bending energy\cite{Helfrich}.
The differential geometry tells us that the mean curvature $H({\bf r})$ can be obtained by a variation of ${\cal S}$ with respect to an infinitesimal displacement of the membrane surface in its perpendicular direction.  
Therefore, the mean curvature $H$ and the Helfrichfs bending energy $F_{\rm b}$ of the vesicle surface are described in terms of $\psi$ as\cite{Du_2} 
\begin{eqnarray}
H({\bf r }) &=&  -\psi({\bf r}) + \psi({\bf r})^{3} - \epsilon^{2} \nabla^{2} \psi({\bf r}), \\
\label{bending}
F_{\rm b}       &=&  \frac{\kappa}{2} \int \left( H({\bf r}) - \sqrt{2} H_{0} ( 1-\psi({\bf r})^{2} )   \right)^{2} d{\bf r},
\end{eqnarray}
where $H_{0}$ is a spontaneous curvature and $\kappa$ is a bending modulus of the vesicle\cite{Helfrich}.
  
The equilibrium shape of a vesicle without polymers is uniquely determined by minimizing the Helfrich's bending energy eq.(\ref{bending}), under the conditions of fixed surface area ${\cal S}_{0}$ and fixed enclosed volume ${\cal V}_{0}$ of the vesicle.
Therefore, we should minimize $F_{\rm PF}$  described by
\begin{equation}
F_{\rm PF} = F_{\rm b} + \sigma \left( {\cal S} - {\cal S}_{0}  \right) + \gamma \left( {\cal V} - {\cal V}_{0} \right),
\end{equation} 
where $\sigma$ and $\gamma$ are Lagrange multipliers that correspond to the surface tension of the vesicle and the osmotic pressure difference between inside and outside regions of the vesicle, respectively.  
The total enclosed volume ${\cal V}$ is defined by ${\cal V}=\int_{\psi>0} d{\bf r}$ because the region with positive value of $\psi$ represents the inner region of the vesicle\cite{VolumeConstraint}. 

While the vesicle is described by PFT, the equilibrium distribution and conformation of the enclosed polymer chains and the solvent are described by SCFT\cite{Kawakatsu,Fredrickson}.  
In the SCFT, each polymer chain is modeled by a linear flexible string made by a sequence of $N$ segments.  The spatial distribution of the polymer segments $\phi_{\rm p}({\bf r})$ is evaluated using a representative chain in a mean field $V_{\rm p}({\bf r})$. 
The probability distribution of the conformation of this representative chain is described by a path-integral $Q_{\rm p}(0,{\bf r}_{0};N,{\bf r}_{N})$ which is defined as the sum of the statistical weights for all the conformations of the representative chain with 
the 0-th segment and the $N$-th segment fixed at positions ${\bf r}_{0}$ and ${\bf r}_{N}$, respectively.  
This path integral is obtained by solving the following Edwards equation
\begin{equation}
\label{PATH}
\frac{\partial Q_{\rm p}(0,{\bf r}_{0};s,{\bf r})}{\partial s} = \left( \frac{b^{2}}{6} \nabla^{2} - \frac{1}{k_{\rm B} T} V_{\rm p}({\bf r}) \right) Q_{\rm p}(0,{\bf r}_{0};s,{\bf r}),
\end{equation} 
where $b$ is the size of the segment, $k_{\rm B }$ the Boltzmann constant and $T$ the temperature.  
The confinement of the polymer is realized by solving eq.(\ref{PATH}) only for the region with $\psi > 0$, where the vesicle surface is treated as a Dirichlet boundary for the path integral $Q_{\rm p}(0,{\bf r}_{0};N,{\bf r}_{N})$.  
This Dirichlet boundary condition is required to estimate the correct conformational entropy of the polymer chains\cite{DiMarzio}.

Once the mean field $V_{\rm p}({\bf r})$, the path integral $Q_{\rm p}(0,{\bf r}_{0};N,{\bf r}_{N})$ and the segment distribution $\phi_{\rm p}({\bf r})$ are obtained, we can evaluate the free energy of the equilibrium state of the polymer solution, $F_{\rm SCF}$, using the formula\cite{Kawakatsu,Fredrickson}
\begin{eqnarray}
\label{SCF_1}
F_{\rm SCF} &=& -M_{\rm p} k_{\rm B}T \int \int_{\psi>0} \ln Q_{\rm p}(0,{\bf r}_{0};N,{\bf r}_{N}) d{\bf r}_{0} d{\bf r}_{N} \nonumber \\
            &-& \int_{\psi>0} \phi_{\rm p}({\bf r})V_{\rm p}({\bf r}) d{\bf r} + \chi \int_{\psi > 0} \phi_{\rm p}({\bf r}) (1-\phi_{\rm p}({\bf r}) ) d{\bf r} \nonumber \\
			&+& k_{\rm B}T \int (1-\phi_{\rm p}({\bf r})) \ln (1-\phi_{\rm p}({\bf r}) ) d{\bf r}, 
\end{eqnarray}
where $M_{\rm p}$ is the total number of polymer chains and $\chi$ is the Flory-Huggins interaction parameter between the polymer segment and the solvent.  
The sum of the first and the second terms on the right-hand side of eq.(\ref{SCF_1}) represents the contributions from the entropy of the polymers, 
the third term is the contribution from the interaction energy between the polymer and the solvents,
and the last term is the contribution from the translational entropy of the solvents.  
In eq.(\ref{SCF_1}), the integrations are taken over the inside region of the vesicle ($\psi > 0$) except for the last term.
In order to minimize the total free energy, we introduce an artificial relaxation dynamics for the phase field $\psi$ given by
\begin{equation}
\label{TIME}
\frac{\partial \psi({\bf r})}{\partial t} = - \frac{\delta \left( F_{\rm PF} + F_{\rm SCF}  \right)}{\delta \psi({\bf r})},
\end{equation}
where $t$ is an artificial time.  
Equation (\ref{TIME}) is used only for obtaining the equilibrium structures of the system.  
A realistic dynamics is proposed in our separate paper\cite{Oya_2}

In the case of a vesicle neither with the spontaneous curvature nor with the enclosed polymers, its equilibrium shape is uniquely determined by a parameter named reduced volume $\upsilon$, 
which is defined by the ratio between the enclosed volume ${\cal V}_{0}$ of the vesicle and the volume of a sphere with the same total surface area ${\cal S}_{0}$ as the target vesicle, 
{\it i.e.} $\upsilon={\cal V}_{0}/\frac{4 \pi}{3}\left( \frac{{\cal S}_{0}}{4 \pi} \right)^{\frac{3}{2}}$.  
In the present study, we set $\upsilon=1/\sqrt{2}$ by fixing ${\cal V}_{0}$ and ${\cal S}_{0}$.   This value $v = 1/\sqrt{2}$ corresponds to a vesicle made by a fusion of two spherical vesicles with the same size.
We assume that the mean volume fraction of the polymers of the enclosed polymer solution $\rho$ is fixed to $\rho = 0.1$. 
To solve the phase field and the path integral, we use a cylindrical coordinate system with $256 {\times} 80$ mesh points in the axial and radial directions, respectively, with a mesh width $\bigtriangleup x = 0.5$.  The total chain length $N$ is discredited with a mesh width $\bigtriangleup s = 0.1$.
In the following, the simulation results are shown for two cases for the vesicles with and without the spontaneous curvature $H_{0}$.

\begin{figure}[t]
   \begin{center}
   \includegraphics[width=70mm]{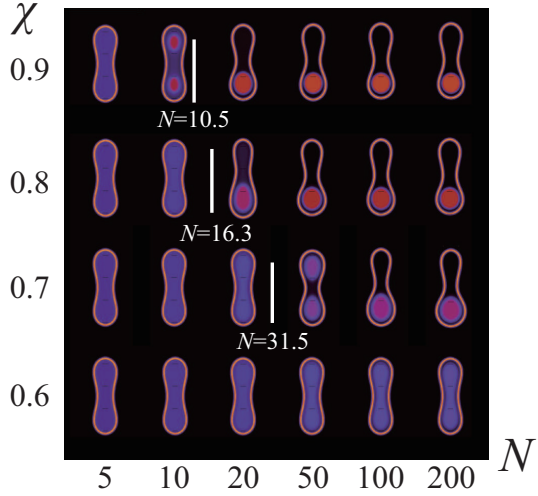}
\caption{\label{PR_DIAGRAM} 
Equilibrium shapes of the vesicle and the equilibrium distributions of the polymers for various values of the $\chi$-parameter and the chain length $N$.  This is the case without the spontaneous curvature ($H_0 = 0$), and the reduced volume and the polymer volume fractions are kept constant at $v = 1/\sqrt{2}$ and $\rho = 0.1$.
The white lines are the boundaries between the uniform states and the phase-separated states of the polymer solution determined by the Flory-Huggins theory for bulk systems.}
  \end{center}
\end{figure}

First, we discuss a polymer-containing vesicle without the spontaneous curvature.  
In fig.\ref{PR_DIAGRAM}, we show equilibrium structures of the vesicle and the polymers for various values of the interaction parameter $\chi$ and the polymer length $N$. 
When the value of $\chi N$ is small, the polymer solution is in a uniform state and the polymers distribute homogeneously inside the vesicle.  In this case, the vesicle keeps the symmetric prolate shape as in the case without the enclosed polymers.  
On the other hand, when $\chi N$ becomes larger, an asymmetric pear shape of the vesicle appears due to the phase separated structures of the polymer solution. 

Here we focus on a transition point between the uniform state and the phase separated state of the polymer solution.  
For a bulk polymer solution without any confinements, the phase separation between the polymer and the solvent takes place at the coexistence curve of the well-known Flory-Huggins free energy\cite{Kawakatsu}.
In fig.\ref{PR_DIAGRAM}, we show the phase transition points obtained by the Flory-Huggins theory by white lines, which are close to those of the simulations but with a slight difference. Such a difference between the simulation and the theory is attributed to the existence of the depletion layers.  
Near the vesicle surface, there is a region where the polymers cannot exist due to the excluded volume interaction with the membrane.  
Such a region is called a "depletion layer".  
The existence of such a depletion layer leads to an increase in the polymer-concentration effectively, leading to a phase separation of the confined polymer solution even if it does not in the bulk system\cite{Lipowsky}.
\begin{figure}[t]
   \begin{center}
   \includegraphics[width=70mm]{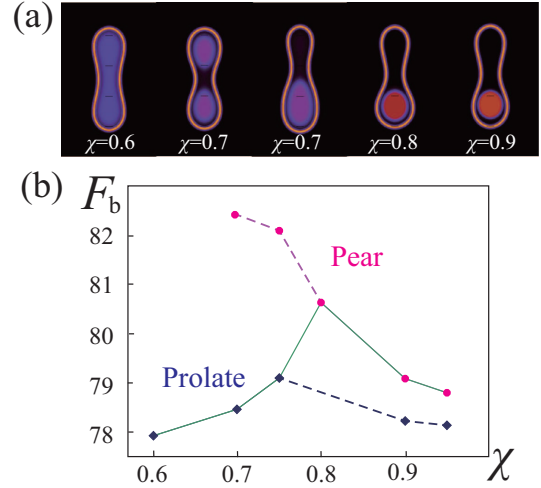}
\caption{\label{PR_PE} 
(a) The stable (or metastable) structures of the vesicle and the polymers are shown for various values of the interaction parameter $\chi$, where $N = 50$ and the other parameters are the same as those in fig.\ref{PR_DIAGRAM}.  The left-side 2 shapes are prolate shapes while the right-side 3 shapes are pear shapes, respectively. 
(b) The bending elastic energies of the prolate shapes($\blacklozenge$) and the pear shapes($\bullet$) are shown as functions of $\chi$.  The stable branch that minimizes the total free energy is indicated by the green line, which shows that the pear shape at $\chi = 0.7$ is metastable.}
  \end{center}
\end{figure}

When $\chi N$ becomes large, the vesicle changes its shape due to the aggregations of the polymers.  
In fig.\ref{PR_PE}(a), two types of equilibrium structures of the polymer-containing vesicle, {\it i.e.} prolate and pear shapes, are shown for $N=50$.  
When $\chi$ becomes larger than 0.8, the polymer segregates in one side of the vesicle and forms a single polymer-rich domain.  
Accordingly, the vesicle changes its shape to an asymmetric pear shape induced by this domain structure of the polymers.  
We also recognize that the transition between the prolate shape and the pear shape is first order, because there are metastable structures such as the pear shape for $\chi = 0.7$ in figs.\ref{PR_PE}.
\begin{figure}[t]
   \begin{center}
   \includegraphics[width=85mm]{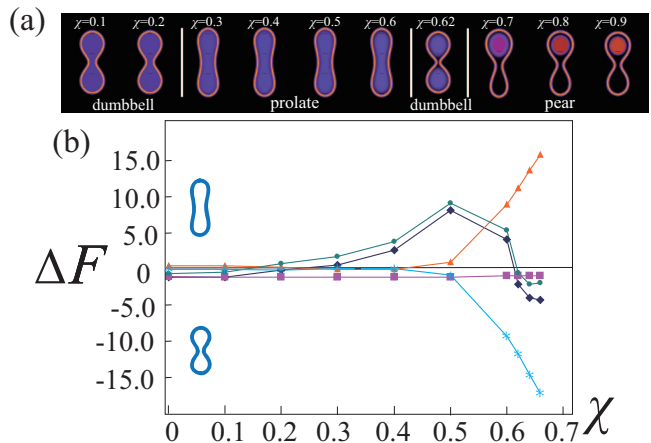}
\caption{\label{PR_SP} 
(a)The equilibrium structures of polymer-containing vesicles that have a non-zero spontaneous curvature $H_0$, and 
(b)the differences between the contributions to the free energy between the dumbbell and the prolate shapes for the case with $N=100$ and $\rho = 0.1$.  
In fig.(b), total free energy ($\blacklozenge$), sum of conformational and translational entropies of the polymers ($\bullet$), translational entropy of the solvents ($\blacktriangle$), bending energy of the membrane ($\blacksquare$), and interaction energy between the polymers and solvents ($\ast$) are shown. 
The regions with positive (negative) values of $\bigtriangleup F$ mean that the prolate shape (dumbbell shape) is more stable compared with the dumbbell shape (prolate shape).}
  \end{center}
\end{figure} 

Next, we discuss the case with non-zero spontaneous curvature $H_{0}$.  
It is known that the spontaneous curvature is essential for the deformation of a multicomponent vesicle or monolayer membrane such as microemulsions\cite{Seifert}.  
We chose the value of the spontaneous curvature as $H_{0}=\sqrt{ {\cal S}/4 \pi }$.  
In fig.\ref{PR_SP}(a), equilibrium shapes of such polymer-containing vesicles with  non-zero $H_{0}$ are shown.  
With increasing $\chi$, the vesicle changes its equilibrium shapes from dumbbell shapes($0<\chi<0.2$) to prolate shapes($0.3<\chi<0.62$), then to dumbbell shapes again($0.62<\chi<0.66$), and finally to pear shapes($\chi>0.66$).

When the shape change in fig.\ref{PR_SP}(a) takes place, each term in the free energy in eq.(\ref{SCF_1}) changes as is shown in fig.\ref{PR_SP}(b). 
Around $\chi = 0.5$, we observe an abrupt change in the behavior of the polymer entropy ($\bullet$) as well as the total free energy($\blacklozenge$).
For $\chi < 0.5$, the polymer entropy dominates the total free energy, while for $\chi > 0.5$ the translational entropy of the solvent($\blacktriangle$) and that from the interaction energy between the polymer and the solvent($\ast$) are dominant.  We can explain this behavior in terms of the gradual shrinkage of the polymer coil according to the increase in $\chi$ for $\chi < 0.5$ and a sudden collapse of the polymer coil at $\chi = 0.5$ (coil-globule transition).  

For very small $\chi$($\chi < 0.2$), the polymer chains are well swollen by the solvent so that their gyration radii are larger than the radius of the vesicle.  As the polymer volume fraction ($\rho = 0.1$) is not small enough, the polymers fill inside the vesicle almost uniformly and depletion layers are not formed.  In such a case, the polymers impose a uniform and isotropic pressure to the vesicle just like the solvent, which leads to the same equilibrium shape as in the case without the enclosed polymers, {\it i.e.} the dumbbell shape. 

When $\chi$ becomes larger but still smaller than 0.5, the polymer chains gradually shrink.  When the size of the polymer chains becomes comparable or smaller than the radius of the vesicle, a thin depletion layer is formed near the vesicle surface, which constrains the translational entropy of the polymers. 
For the case of the dumbbell shaped vesicle, the polymers spread inside the 2 spherical compartments of the vesicle, leading to a strong constraint of the centers of mass of the polymers into the central regions of the 2 compartments.
On the other hand, for the case of the prolate shaped vesicle, the centers of mass of the polymer chains are constrained into the narrow cylindrical region along the axis of revolution of the prolate shape.  As a result, the polymer chains have more freedom in the prolate-shaped vesicle than in the dumbbell-shaped vesicle, leading to the increased preference of the prolate shape when $\chi$ is increased. 

The role of the depletion layer suddenly changes at $\chi = 0.5$ where the polymer chains drastically shrink due to the coil-globule transition.  For $\chi > 0.5$, the polymer chains can be regarded as small particles, which have negligibly small conformational entropy.  Therefore, the inner pressure of the vesicle is mainly arising from the translational entropy of the polymers, which is strongly dependent on the volume of the depletion layer.  Here, the dumbbell shape has a smaller volume of the depletion layer than that of the prolate shape because of a large overlap between different sections of the depletion layer around the neck region.  This is the reason why the re-entrant behavior of the dumbbell shape for $0.62 < \chi < 0.66$ takes place.

A further increase in the $\chi$-parameter leads to a complete separation between the polymer-rich phase and the solvent-rich phase, which induces another transition to the asymmetric pear shape as in the case for the vesicle without spontaneous curvature (fig.\ref{PR_DIAGRAM}).

Recently, Terasawa, {\it et al.} reported experimentally that the vesicles can undergo a fission in order to reduce the depletion layers\cite{Terasawa}, which is consistent with our finding of the transition from the prolate shape to the dumbbell shape around $\chi = 0.62$.  

In conclusion, we studied polymer-containing vesicles by coupling the phase field theory and the self-consistent field theory.  
We focused on the deformations of the vesicle induced by the phase separation between the polymers and the solvents inside the vesicle.  
We found that the domain structures of the polymers lead to a first order phase transition between a symmetric prolate shape and an asymmetric pear shape of the vesicle.  
Moreover, vesicles with non-zero spontaneous curvatures show a large variety of shapes, such as dumbbell, prolate and pear shapes.  
These shape transitions are induced by the depletion layers for the polymers inside the vesicle.  

{\bf Acknowledgements}
The authors thank M.Imai, T.Taniguchi, T.Murashima, Y.Sakuma for valuable discussions.
The present work is supported by the Grant-in-Aid for Scientific Research from the Ministry of Education, Culture, Sports, Science, and Technology of Japan, and by the Global-COE program at Tohoku University.

\end{document}